\documentclass[a4paper]{jpconf}
\usepackage{graphicx}
\usepackage{amsmath}
\usepackage{amssymb}

\begin{document}

\title{Identified particle production in pp collisions at $\mathbf{\sqrt{\mathit{s}}=7}$ and 13 TeV measured with ALICE}

\author{R. Derradi de Souza (for the ALICE Collaboration)}

\address{Instituto de F\'isica Gleb Wataghin, UNICAMP, 13083-859 Campinas SP, Brazil}

\ead{rderradi@ifi.unicamp.br}

\begin{abstract}
Proton-proton (pp) collisions have been used extensively as a reference for the study of interactions of larger colliding systems at the LHC.
Recent measurements performed in high-multiplicity pp and proton-lead (p-Pb) collisions have shown features that are reminiscent of those observed in lead-lead (Pb-Pb) collisions.
In this context, the study of identified particle spectra and yields as a function of multiplicity is a key tool for the understanding of similarities and differences between small and large systems.
We report on the production of pions, kaons, protons, $K^{0}_{\rm S}$, $\Lambda$, $\Xi$, $\Omega$ and $K^{*0}$ as a function of multiplicity in pp collisions at $\sqrt{s}=$ 7 TeV measured with the ALICE experiment.
The work presented here represents the most comprehensive set of results on identified particle production in pp collisions at the LHC.
Spectral shapes, studied both for individual particles and via particle ratios as a function of $p_{\rm T}$, exhibit an evolution with charged particle multiplicity that is similar to the one observed in larger systems.
In addition, results on the production of light flavour hadrons in pp collisions at $\sqrt{s}=$ 13 TeV, the highest centre-of-mass energy ever reached in the laboratory, are also presented and compared with previous, lower energy results.
\end{abstract}

\section{Introduction}
Recent results of proton-proton (pp) and proton-lead (p-Pb) collisions at the LHC energies have shown features that are similar to those observed in collisions of larger systems such as lead-lead (Pb-Pb).
Notably, these findings include the observation of the double ridge structure in two-particle correlation studies \cite{Khachatryan:2010gv,Abelev:2012ola,Aad:2015gqa} and
the hardening of the transverse momentum spectra of identified particles \cite{Abelev:2013haa} (see also Ref.\ \cite{Loizides:2016tew} for a short summary of experimental results from small system measurements at the LHC).
Such results have motivated several theoretical investigations \cite{Dusling:2013qoz,Ortiz:2013yxa,Bozek:2013uha} as well as an effort on the experimental side in order to provide further and more comprehensive data from small systems collisions.
In this context, studying identified particle production systematically as a function of charged particle multiplicity in pp may provide further insights into the dynamics of small systems.

\section{Analysis Description}
Light flavor hadrons ($\pi$, K, p) are identified in the ALICE experiment \cite{Aamodt:2008zz} using combined information from ITS, TPC and TOF detectors \cite{Adam:2015qaa}.
An extended $p_{\rm T}$ range (up to $\sim$20 GeV/$c$) is achieved with statistical particle identification via the relativistic rise of the ${\rm d}E/{\rm d}x$ in the TPC \cite{Abelev:2014ffa}.
The measurements of strange and multi-strange hadrons ($K^{0}_{S}$, $\Lambda$, $\Xi$ and $\Omega$) are performed via the invariant mass distribution obtained from their respective decay products in the following channels (branching ratios): $K^{0}_{S} \rightarrow \pi\pi$ (69.2\%), $\Lambda \rightarrow p \pi$ (63.9\%), $\Xi \rightarrow \Lambda \pi$ (99.9\%) and $\Omega \rightarrow \Lambda K $ (67.8\%).
A set of topological cuts is applied to the candidates so that those which do not fit into the expected decay are eliminated and the extraction of the raw yield is performed via bin counting technique as done in previous work \cite{Aamodt:2011zza}.
Finally, short lived hadronic resonances ($K^{*0}$, $\bar{K}^{*0}$) are also measured via the invariant mass reconstruction of their decay products ($K\pi$).
In this case, combinatorial background is estimated using a mixed-event technique and subtracted from the original distribution. 
The raw yield is then extracted from a fit of the signal peak \cite{Abelev:2012hy}.
In all analyses, the transverse momentum spectra are corrected for acceptance and efficiency, which are determined with Monte Carlo simulations of the ALICE detector response.

Multiplicity classes are defined based on the total charge deposited in the V0A and V0C detectors located at forward ($2.8<\eta<5.1$) and backward ($-3.7<\eta<-1.7$) pseudorapidity regions, respectively.
The average charged particle density, $\langle{\rm d}N_{\rm ch}/{\rm d}\eta\rangle$, is estimated within each multiplicity class by the average number of charged tracks in the region $|\eta|<0.5$.

\section{Results}
\begin{figure}[h]
\includegraphics[width=32pc]{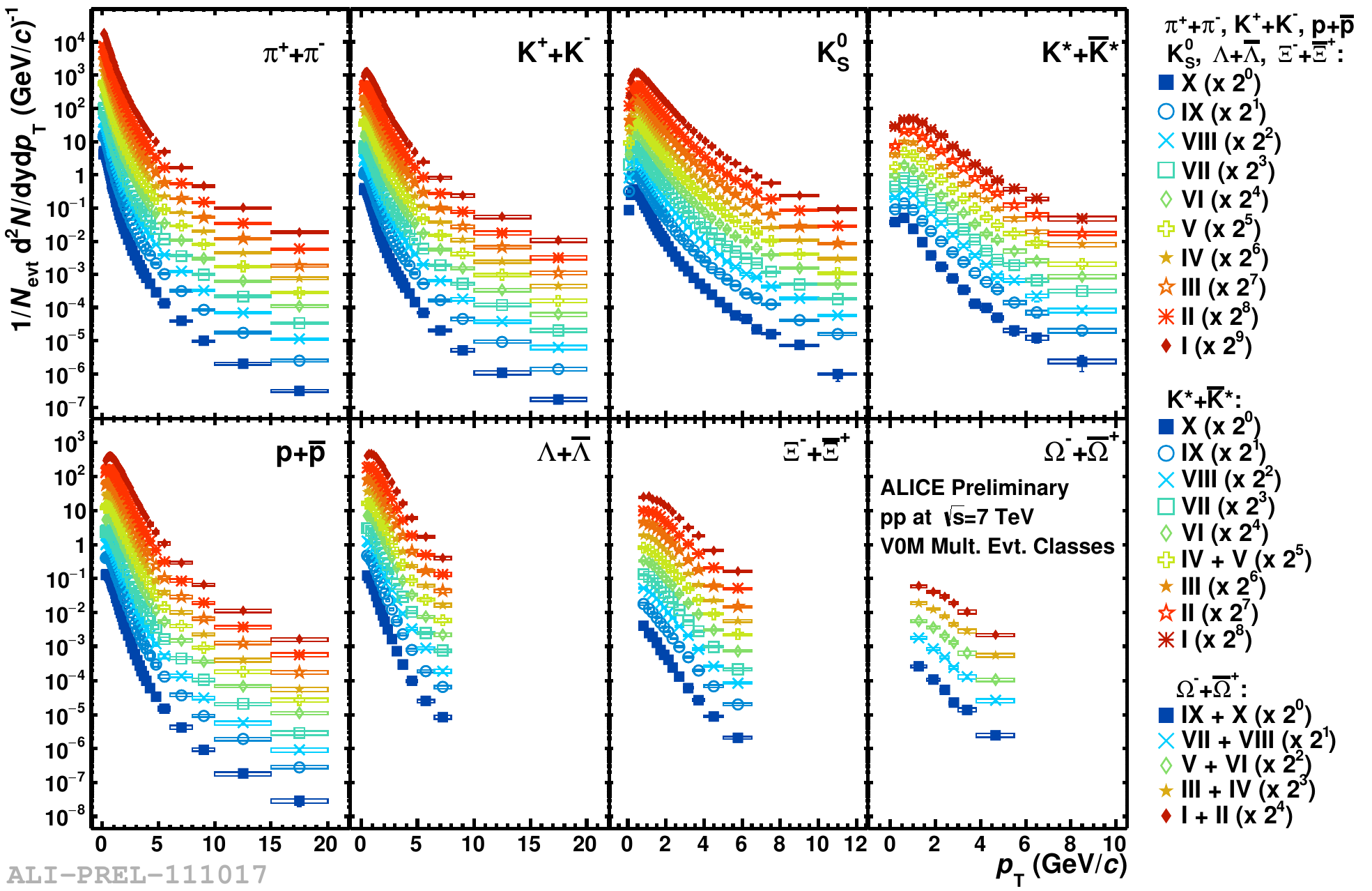}
\caption{\label{fig:pT_spectra}$p_{\rm T}$-spectra of identified particles measured in pp collisions at $\sqrt{s}=$ 7 TeV for several multiplicity classes, labeled by Roman numerals from high (class I) to low (class X) multiplicities.}
\end{figure}

\begin{figure}[h]
\includegraphics[width=26pc]{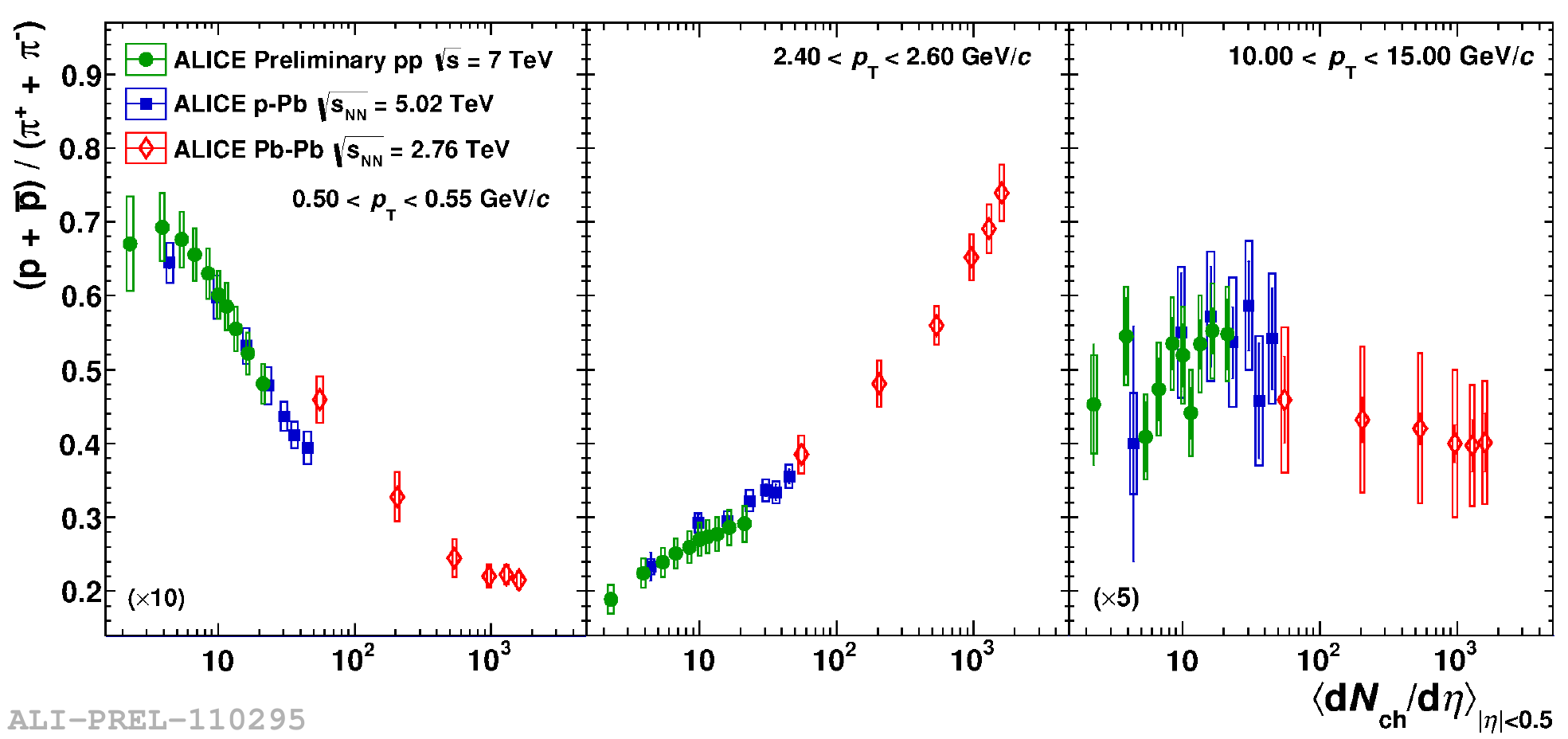}
\hspace{1pc}
\begin{minipage}[b]{10.5pc}
\caption{\label{fig:ProtonOverPion_vs_Nch}Proton-to-pion ratio at low (left panel), intermediate (middle panel) and high (right panel) transverse momentum as a function of the average charged particle density at midrapidity for pp, p-Pb and Pb-Pb.}
\end{minipage}
\end{figure}

Figure \ref{fig:pT_spectra} shows the final transverse momentum spectra of the identified hadrons measured in pp collisions at 7 TeV for several event multiplicity classes (indicated by Roman numerals).
The average charged particle densities in the highest (class I) and lowest (class X) multiplicity classes correspond to approximately 3.5 and 0.4 times the value in the integrated sample ($\langle{\rm d}N_{\rm ch}/{\rm d}\eta\rangle \approx 6.0$), respectively.
A clear hardening of the $p_{\rm T}$ spectra is observed for all species from low to high multiplicity events.
Moreover, this effect seems to be slightly more pronounced for baryons than for mesons.
In fact, the baryon-to-meson ratio in pp collisions has been recently shown to follow a similar trend between lowest and highest multiplicity classes as observed in p-Pb and Pb-Pb collisions \cite{Bianchi:2016szl}.
In order to further explore the evolution of the different physics mechanisms related to baryon and meson production at different regions of the $p_{\rm T}$-spectra, the ratio of proton-to-pion was studied at low-, mid- and high-$p_{\rm T}$ regions as a function of $\langle{\rm d}N_{\rm ch}/{\rm d}\eta\rangle$, as shown in Fig.\ \ref{fig:ProtonOverPion_vs_Nch}.
%
%
%
The continuity observed across the various collision systems is remarkable for all three $p_{\rm T}$ regions.
The smooth evolution of particle ratios suggests the similarity of the underlying mechanisms despite the differences in the initial colliding systems.
%
Comparisons with predictions from QCD-inspired models such as PYTHIA8 \cite{Sjostrand:2007gs}, DIPSY \cite{Flensburg:2011kk,Bierlich:2014xba,Bierlich:2015rha} and HERWIG7 \cite{Bahr:2008pv,Bellm:2015jjp} have also been performed.
\begin{figure}[h]
\includegraphics[width=15pc]{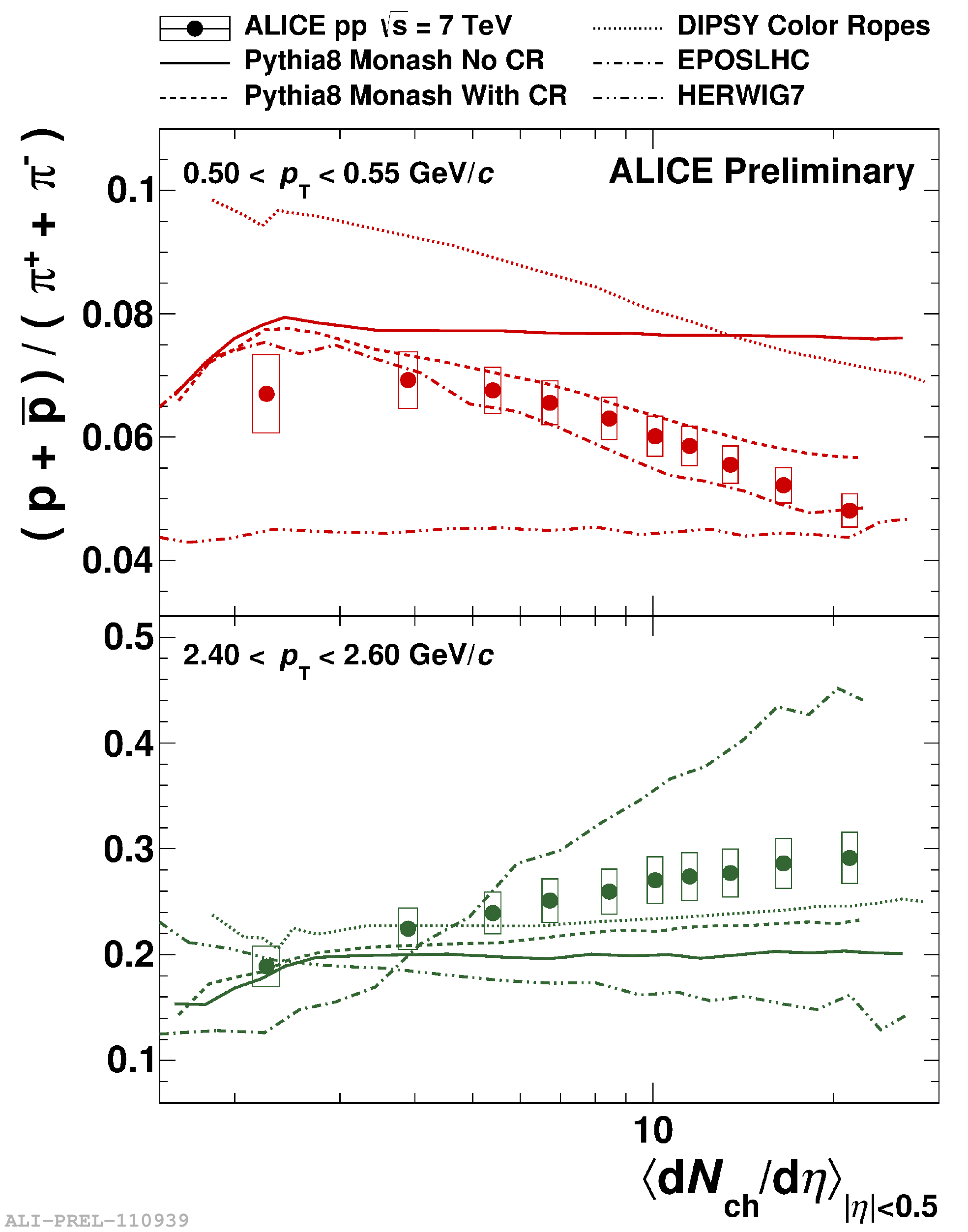}
\hspace{2pc}
\begin{minipage}[b]{13pc}
\caption{\label{fig:ProtonOverPion_vs_Nch_MC}Proton-to-pion ratio at low (top panel) and intermediate (bottom panel) transverse momentum as a function of the average charged particle density at midrapidity for pp collisions at $\sqrt{s}=$ 7 TeV (symbols) compared to model predictions (lines).\\}
\end{minipage}
\end{figure}
The color reconnection mechanism in PYTHIA8 as well as the color ropes implementation in DIPSY seem to qualitatively reproduce the general trend observed at low and intermediate $p_{\rm T}$ (see Fig.\ \ref{fig:ProtonOverPion_vs_Nch_MC}).
An alternative description is provided by EPOS LHC \cite{Pierog:2013ria}, which implements collective radial flow using hydrodynamics, and also follows the data qualitatively.
However, none of the models tested were able to quantitatively describe the data for the two $p_{\rm T}$ regions simultaneously.

\begin{figure}[h]
\begin{minipage}{14pc}
\includegraphics[width=13pc]{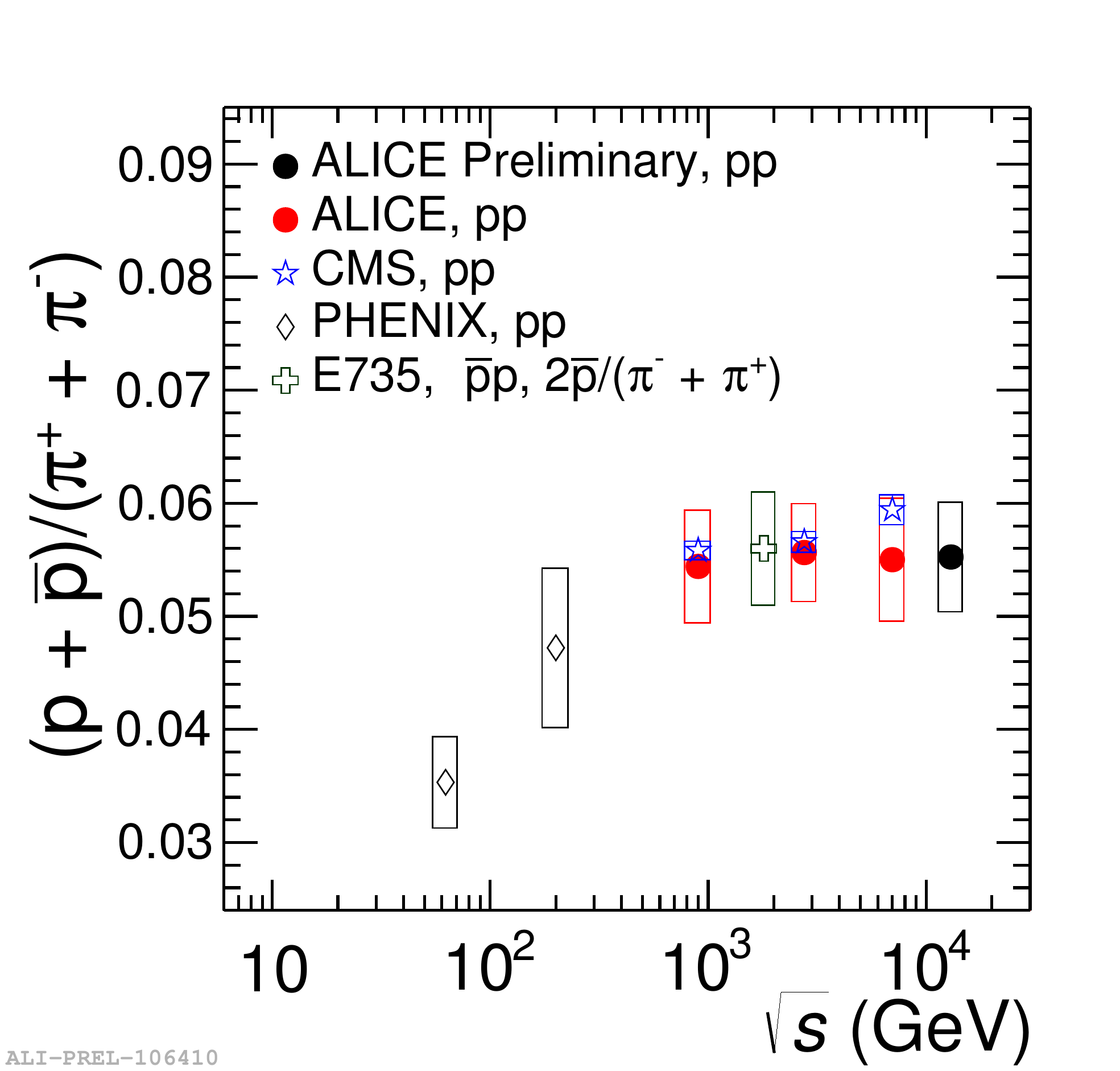}
\caption{\label{fig:ProtonOverPion_vs_sqrts}Proton-to-pion ratios as a function of $\sqrt{s}$.}
\end{minipage}
\hspace{2pc}
\begin{minipage}{17pc}
\includegraphics[width=16pc]{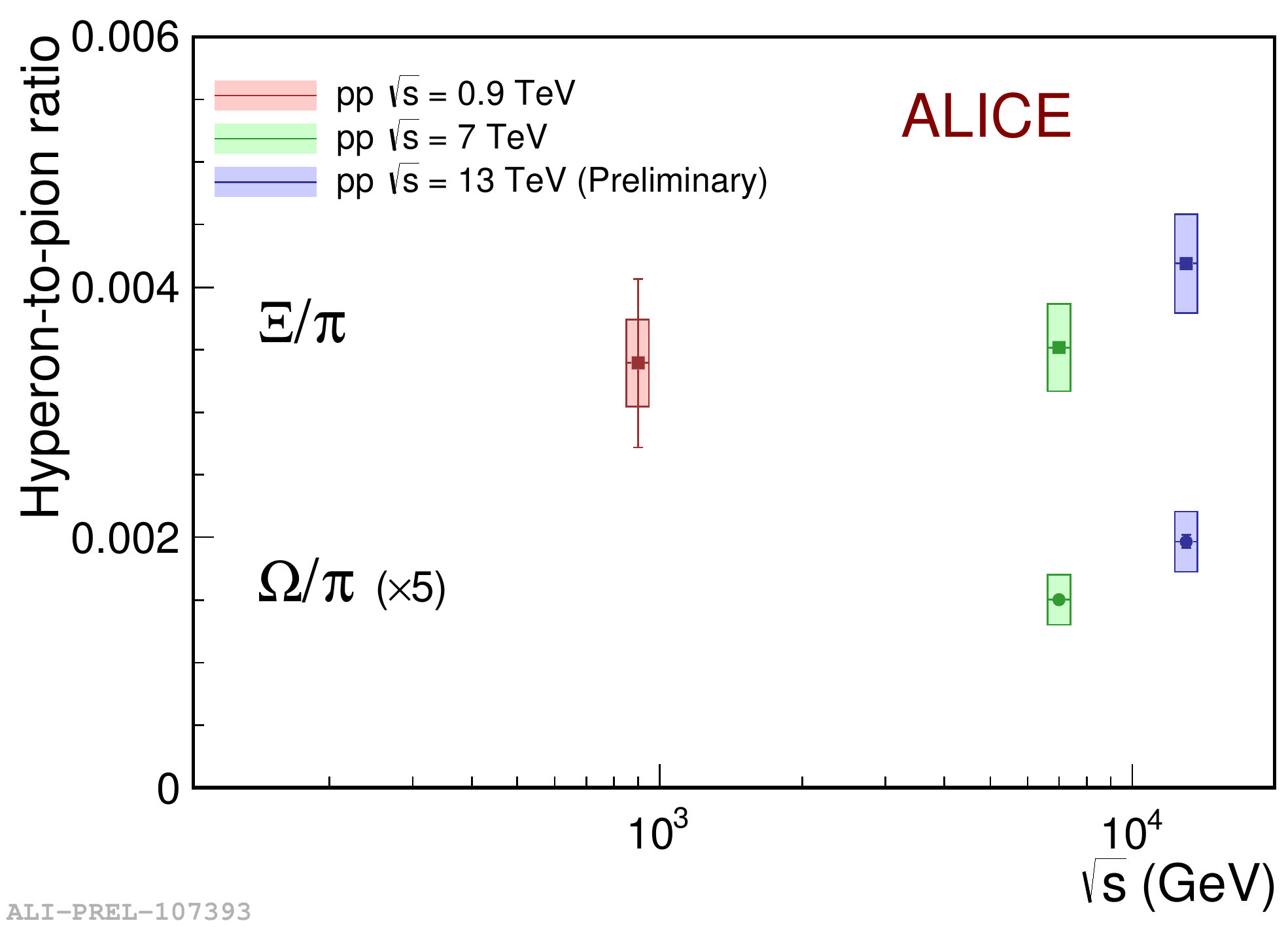}
\caption{\label{fig:HyperonOverPion_vs_sqrts}Hyperon-to-pion ratios as a function of $\sqrt{s}$ at LHC energies.}
\end{minipage} 
\end{figure}

The evolution of particle ratios with the collision energy is another important aspect which may reveal essential details concerning the underlying particle production mechanisms.
Figures \ref{fig:ProtonOverPion_vs_sqrts} and \ref{fig:HyperonOverPion_vs_sqrts} show the ratios with respect to pions for $p_{\rm T}$-integrated yields of protons and hyperons, respectively, as a function of the collision energy.
The yields of individual species are obtained as done in previous analysis \cite{Adam:2015qaa,Abelev:2012jp}.
The values observed for proton-to-pion ratio seem to saturate at the LHC energies, with the latest point added at 13 TeV following the trend of previous measurements.
On the other hand, the hyperon-to-pion ratios present a slight increase at 13 TeV with respect to lower collision energies.

\section{Summary and Outlook}
A comprehensive set of measurements of identified hadron production in pp collisions is presented and a detailed investigation of the underlying particle production mechanisms is currently ongoing.
Furthermore, the increase as a function of collision energy of the hyperon-to-pion ratios at 13 TeV may also be related to the enhanced production of strange and multi-strange hadrons as a function of event multiplicity observed in pp collisions at 7 TeV \cite{Adam:2016emw}, as the mean charged particle multiplicity density also increases with collision energy \cite{Adam:2015pza}.
The study of the multiplicity dependence of particle ratios at 13 TeV in combination with the reported results at 7 TeV will provide crucial input to disentangle the dependence of particle production with collision energy and event multiplicity.

This work was supported by grant 2014/09167-8, S\~ao Paulo Research Foundation (FAPESP).

\section*{References}

\bibliographystyle{iopart-num}
\bibliography{references}

\end{document}